# Density-independent plasmons for terahertz-stable topological metamaterials

Jianfeng Wang[1]*, Xuelei Sui[1,2], Wenhui Duan[2], Feng Liu[3] and Bing Huang[1,4]*

[1] *Beijing Computational Science Research Center, Beijing 100193, China*
[2] *Department of Physics and State Key Laboratory of Low-Dimensional Quantum Physics, Tsinghua University, Beijing 100084, China*
[3] *Department of Materials Science and Engineering, University of Utah, Salt Lake City, Utah 84112, USA*
[4] *Department of Physics, Beijing Normal University, Beijing 100875, China*

*Emails: bing.huang@csrc.ac.cn; wangjf@csrc.ac.cn

**Abstract**

**To efficiently integrate cutting-edge terahertz technology into compact devices, the highly confined terahertz plasmons are attracting intensive attentions. Compared to plasmons at visible frequencies in metals, terahertz plasmons, typically in lightly doped semiconductors or graphene, are sensitive to carrier density ($n$) and thus have an easy tunability, which, however, leads to unstable or imprecise terahertz spectra. By deriving a simplified but universal form of plasmon frequencies, here we reveal a unified mechanism for generating unusual $n$-independent plasmons (DIPs) in all topological states with different dimensions. Remarkably, we predict that terahertz DIPs can be excited in 2D nodal-line and 1D nodal-point systems, confirmed by the first-principles calculations on almost all existing topological semimetals with diverse lattice symmetries. Besides of $n$ independence, the feature of Fermi-velocity and degeneracy-factor dependences in DIPs can be applied to design topological superlattice and multi-walled carbon nanotube metamaterials for broadband terahertz spectroscopy and quantized terahertz plasmons, respectively. Surprisingly, high spatial confinement and quality factor, also insensitive to $n$, can be simultaneously achieved in these terahertz DIPs. Our findings pave the way to developing topological plasmonic devices for stable terahertz applications.**

Bridging the gap between microwave and infrared regimes, terahertz radiation promises many cutting-edge applications in radar, imaging, biosensing, nondestructive evaluation and ultrahigh-speed communications[1,2]. While realizing compact terahertz integrated circuits is a big challenge, terahertz plasmons, collective oscillations of electrons at terahertz frequency, provide a revolutionary way to effectively reduce the sizes of terahertz devices down to sub-wavelength scales[3–8]. To achieve highly confined terahertz plasmons, the extensive research has been devoted to various metamaterials, including spoof plasmon polaritons in structured metal surfaces[7–10], terahertz plasmons in lightly doped semiconductors[2,11–13] and recently developed graphene plasmons[14–16]. Compared to plasmons at visible frequency in metals, terahertz plasmons, *e.g.*, in semiconductors and graphene, are quite sensitive to the oscillation of carrier density ($n$)[13–18], as a low $n$ can be greatly changed by the defects[17], thermal fluctuation[2,16,19], charge inhomogeneity[20], electrical gating[14,16,18], optical excitations[21–23] or charge transfer at interface (Fig. 1a). Consequently, their fundamental properties, such as resonance frequency, confinement and loss of terahertz plasmons[2,16–18], will be largely affected by the surrounding environments. Therefore, the $n$ dependence feature leads to unfavorable terahertz applications, such as low temperature limit, high-quality sample requirements, unstable or imprecise terahertz sources and detection.

It is known that the classical plasmon frequency in conventional electronic gas (EG) has a $n^{1/2}$ dependence, while graphene plasmon shows a weaker $n^{1/4}$ power-law scaling[14,15]. Recently, the linear band structures have been extended to a large number of topological semimetals (TSMs)[24,25], following the fast development of topological matter. Surprisingly, the plasmons with diverse $n$ dependences have been found in these TSMs, even though they have a similar linear band crossing as graphene. For example, the plasmon frequency of 3D Dirac systems shows a $n^{1/3}$ scaling[26], while unconventional $n^0$-dependent plasmons solely in mid-infrared have been found in 1D metallic carbon nanotubes (CNTs)[27] or 3D nodal-surface electrides[28]. Since most of previous theories are system dependent, a unified theory to intuitively understand all these plasmonic behaviors in different electronic systems is still lacking, which significantly prevents the design of superior metamaterials for revolutionary terahertz technology and overcoming the intrinsic terahertz-unstable bottlenecks in conventional plasmonic devices.

In this article, we derive a simplified but universal form of plasmon frequencies at long-wavelength limit that can be applied to understand the collective excitations of all electronic systems with different dimensions. Significantly, a unified mechanism is revealed for generating density-*independent* plasmons (DIPs), which can be excited in some specific topological states. As demonstrated in Fig. 1a, the properties of a DIP, such as its resonance frequency or wavelength, are not affected by the changes of $n$, which can fundamentally overcome intrinsic terahertz-unstable bottlenecks raised by density-dependent plasmons (DDPs) in conventional systems. Importantly, we predict that the terahertz DIPs can be realized in two reduced systems: 2D nodal line and 1D nodal point. Extensive first-principles calculations are employed to confirm the DIP excitations among 22 known 2D nodal-line semimetals (NLSMs) and 1D CNTs. Besides of the $n$ independence, the frequencies of DIPs can be tuned by Fermi velocity, substrate screening and degeneracy factor, revealing that a novel ultrastable terahertz spectrum from narrowband to broadband and a tunable quantization can be achieved in 2D superlattice and 1D multi-walled CNT metamaterials, respectively. Remarkably, stable performance with high spatial confinement and quality factor, critical for device applications, can be simultaneously obtained for terahertz DIPs.

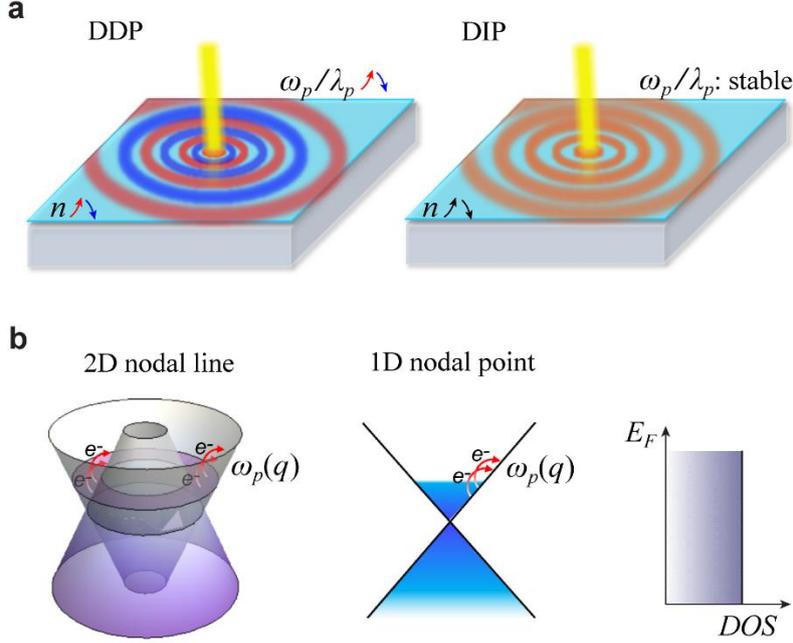

**Figure 1 | DIPs and their realizations. a,** Schematic comparison between *n*-dependent plasmon (DDP) and *n*-independent plasmon (DIP). Concentric red, blue or orange circles illustrate plasmon waves excited by electron systems (represented by the cyan plane). For an excited DDP, its resonance frequency ($\omega_p$) or wavelength ($\lambda_p$) is sensitive to the oscillation of *n*. When increasing (red arrow) or decreasing (blue arrow) *n*, the $\omega_p$ or $\lambda_p$ of a DDP will increase (red circles) or decrease (blue circles) correspondingly. While the properties ($\omega_p$ or $\lambda_p$) of a DIP (orange circles) are stable against the changes of *n*. **b,** Linear band structures of 2D nodal line and 1D nodal point, and their constant DOS versus $E_F$. Collective DIPs are labeled schematically by red arrows.

**A unified mechanism for DIPs**

The plasmon excitation can be determined by the dynamical dielectric function $\epsilon(\boldsymbol{q},\omega) = 1 - V(q)\Pi(\boldsymbol{q},\omega)$, where $V(q)$ is dimension-related Coulomb interaction in the wave vector space, and $\Pi(\boldsymbol{q},\omega)$ is the irreducible polarizability function (see Methods section). As a function of Fermi energy ($E_F$), the key polarizability is usually *n*-dependent. Under RPA and long-wavelength limit, the *D*-dimensional noninteracting irreducible polarizability near $E_F$ can be approximated by Taylor's first-order expansion,

$$\Pi(\boldsymbol{q},\omega) \approx \frac{g}{(2\pi)^D} \int d^D\boldsymbol{k}\, \frac{\partial n_F}{\partial E}\left(\frac{\partial E}{\partial k}\right)^2 \left(\frac{q}{\hbar\omega}\right)^2, \tag{1}$$

where $n_F$ is the Fermi-Dirac distribution function, and $g$ is the degeneracy factor including degeneracies of spin, valley and conducting channel. Around *T*=0 K, we derive, by solving the zeros of dielectric function, a simplified but general form of the plasmon frequency,

$$\omega_p \approx \rho(E_F)^{1/2} v_F V(q)^{1/2} q, \tag{2}$$

where $\rho(E_F)$ is the density of states (DOS) of electrons, and $v_F = \hbar^{-1}(\partial E/\partial k)_F$ is the Fermi velocity. Equation (2), one of our key results, is universal for all dimensional solids, which reveals the most essential elements related to plasmons. Obviously, it shows that, besides of $V(q)$, $\omega_p$ is mostly determined by DOS and $v_F$.

Now we can shed light on the nature of *n* dependence of plasmons for all known systems. For the conventional EGs in metals or doped semiconductors exhibiting a parabolic energy dispersion $E(\bm{k}) = \hbar^2 k^2 / 2m$, their $v_F$ are $E_F$- (or *n*-) dependent. As listed in Table 1, the DOS of conventional EGs is related to their dimensions (see Supplementary Section I). Based on Eq. (2), $\omega_p$ of all conventional EGs has a well-known $n^{1/2}$ power-law scaling (Table 1). While all the TSMs have a linear band dispersion $E(\bm{k}) = \hbar v_F k$, whose $v_F$ is a constant. In terms of the dimensionality of band crossings, TSMs can be classified into nodal point, nodal line and nodal surface[24,25,29]. As demonstrated in Table 1, the $\omega_p$ of TSMs has different scalings, solely dependent on their DOS.

As shown in the right column of Table 1, we have derived rigorously the analytical expressions of $\omega_p$ for all systems with different dimensions (see Methods section and Supplementary Section II), confirming that Eq. (2) could accurately capture the power-law scaling of $\omega_p$, *e.g.*, the $n^{1/4}$ and $n^{1/3}$ DDPs are well reproduced for 2D (*e.g.*, graphene) and 3D nodal-point (*e.g.*, Dirac semimetal) systems, respectively. Interestingly, a plasmon with the same *n*-dependent scaling as graphene also appears in 3D nodal-line systems but it has a different anisotropy. In fact, all the analytical expressions in Table 1 can be obtained from Eq. (2) with only a difference of dimensionless coefficient.

Based on Eq. (2), we can reveal a unified mechanism for realizing a DIP, which needs to meet two general criteria: (i) a constant DOS near $E_F$; (ii) a constant $v_F$. While (ii) can be naturally achieved in the linear band dispersion region of a TSM, (i) is the key criterion for achieving a DIP. After systematic derivations (see Supplementary Sections I and II), we conclude that the constant DOS and the resulting DIPs can solely exist in the following TSM states: 3D nodal surface, 2D nodal line and 1D nodal point, as listed in Table 1. It is noted that the plasmons occurred in wide parabolic quantum wells[30,31] can be independent of the electron numbers (not electron density) under certain conditions, which is fundamentally different from our DIPs. We illustrate the constant DOS of 2D nodal line and 1D nodal point in Fig. 1b. At long-wavelength limit, the excitation process mainly occurs near $E_F$. For 2D nodal line, all states involved are in the vicinity of two rings, and the total numbers of their sum will keep a constant when changing $E_F$ (Fig. 1b). For 1D nodal point, the total states in the excitations are at two points, and their numbers are also unchanged (Fig. 1b). It is noted that all systems are supposed to behave as Fermi liquids, although 1D metallic electrons may also be considered as a Luttinger liquid[32].

As listed in Table 1, the analytical expressions clearly demonstrate the nature of DIPs in 3D nodal surface, 2D nodal line and 1D nodal point. Besides of the *n* independence, their $\omega_p$ are determined by some other physical quantities, *i.e.*, Fermi velocity $v_F$, degeneracy factor $g$ and size of degenerate node ($k_0$ for line node, and $S$ for surface node), giving the tunable factors. Interestingly, the $\omega_p$ of all TSMs is manifestly quantum with an explicit "$\hbar$", in contrast to the classical plasmons of conventional EGs[26]. In addition, the plasmon dispersion is related to the dimension, which is important for the size effect and spatial compression of plasmons. In order to realize the terahertz metamaterials, we focus on 2D nodal-line and 1D nodal-point systems (the high DOS prevents the realization of terahertz $\omega_p$ in 3D nodal surface[28]). The background dielectric constant ($\kappa$) of these two systems is determined by the surrounding media, providing another tunable factor of $\omega_p$.

**Table 1 | Plasmon frequencies in all dimensions at long-wavelength limit**

| | Systems | $\rho$ | $\rho^{1/2} \cdot v_F$ | analytical expression of $\omega_p$ | |
|---|---|---|---|---|---|
| **3D** | conventional EG | $E_F^{1/2}$ ($n^{1/3}$) | $n^{1/2}$ | $\sqrt{\frac{4\pi e^2 n}{\kappa m}},$ | (3.1) |
| | nodal point | $E_F^2$ ($n^{2/3}$) | $n^{1/3}$ | $\sqrt{\frac{e^2 v_F}{\kappa \hbar}}(\frac{32\pi g}{3})^{1/6} n^{1/3},$ | (3.2) |
| | nodal line | $E_F$ ($n^{1/2}$) | $n^{1/4}$ | $\sqrt{\frac{2\pi e^2 v_F}{\kappa \hbar}(1+\sin^2\theta)}(g\pi k_0 n)^{1/4},$ | (3.3) |
| | <span style="color:red">nodal surface</span> | <span style="color:red">$E_F^0$ ($n^0$)</span> | <span style="color:red">$n^0$</span> | $\sqrt{\frac{g e^2 v_F S \cos^2\theta}{\pi^2 \kappa \hbar}},$ | (3.4) |
| **2D** | conventional EG | $E_F^0$ ($n^0$) | $n^{1/2}$ | $\sqrt{\frac{2\pi e^2 n}{\kappa m}} q^{1/2},$ | (3.5) |
| | nodal point | $E_F$ ($n^{1/2}$) | $n^{1/4}$ | $\sqrt{\frac{e^2 v_F}{\kappa \hbar}}(g\pi n)^{1/4} q^{1/2},$ | (3.6) |
| | <span style="color:red">nodal line</span> | <span style="color:red">$E_F^0$ ($n^0$)</span> | <span style="color:red">$n^0$</span> | $\sqrt{\frac{g e^2 v_F k_0}{\kappa \hbar}} q^{1/2},$ | (3.7) |
| **1D** | conventional EG | $E_F^{-1/2}$ ($n^{-1}$) | $n^{1/2}$ | $\sqrt{\frac{2 e^2 n}{\kappa m}} q\sqrt{|\ln(qa)|},$ | (3.8) |
| | <span style="color:red">nodal point</span> | <span style="color:red">$E_F^0$ ($n^0$)</span> | <span style="color:red">$n^0$</span> | $\sqrt{\frac{2g e^2 v_F}{\pi \kappa \hbar}} q\sqrt{|\ln(qa)|}.$ | (3.9) |

Physical quantities: DOS $\rho$, Fermi velocity $v_F$, electron charge $e$, effective mass $m$, background dielectric constant $\kappa$, degeneracy factor $g$, size of line (surface) node $k_0$ ($S$), and lateral confinement size of 1D electron system $a$. See Supplementary Information for the detailed derivations. Systems exhibiting DIPs are highlighted by red colors. All conventional EGs have a parabolic dispersion, while all TSMs have a linear dispersion.

**Terahertz DIPs in 2D NLSMs**

The 2D NLSMs, having a symmetry-protected crossing between conduction and valence bands along a 1D loop in the Brillouin zone (BZ), attract intensive interests, due to their potential applications in quantum devices. Until now, the 2D NLSMs includes at least 22 compounds with a wide range of lattice symmetries, *e.g.*, honeycomb lattice CuSe[33], AgTe[34] and *h*-B$_2$O[35], honeycomb-Kagome lattice Hg$_3$As$_2$[36], honeycomb-triangular lattice Cu$_2$Si[37], Lieb lattice Be$_2$C[38], and tetragonal lattice $X_2Y$ ($X$=Ca,Sr,Ba; $Y$=As,Sb,Bi)[39], etc. These 2D nodal lines are protected by (glide) mirror symmetries. Importantly, all these NLSMs can exhibit the terahertz DIP feature, confirming our unified theory in 2D systems. Here, we take the experimentally synthesized Cu$_2$Si as an example to demonstrate its DIP excitations, leaving the results of other 21 compounds in Supplementary Section IV.

As shown in Fig. 2a, the monolayer Cu$_2$Si is composed of a honeycomb Cu lattice and a triangular Si lattice. All Cu and Si atoms are coplaner and thus a mirror reflection symmetry with respect to $xy$ plane ($M_z$) is kept. First, we illustrate the electronic properties of free-standing Cu$_2$Si. The calculated band structure without spin-orbit coupling (SOC) is shown in Fig. 2b. Two band crossings between one conduction band and two valence bands occur along $\Gamma$–$M$ and $\Gamma$–$K$ lines. Actually, the band crossings take place along two loops in the 2D BZ, as shown in the 2D band plot (Fig. 2c).

Thus, two concentric nodal lines are formed centred around the $\Gamma$ point. With opposite eigenstate parities of $M_z$ for the conduction and two valence bands[37], the two nodal lines are protected by mirror reflection symmetry (Supplementary Fig. S1). Remarkably, the nearly constant DOS maintains over a large energy range near the Fermi level (Fig. 2b), which is the key condition for the formation of DIP excitations in a TSM. After including SOC effect, the degeneracy of 2D nodal line is slightly lifted with the appearance of a negligible gap[37] (Supplementary Fig. S1).

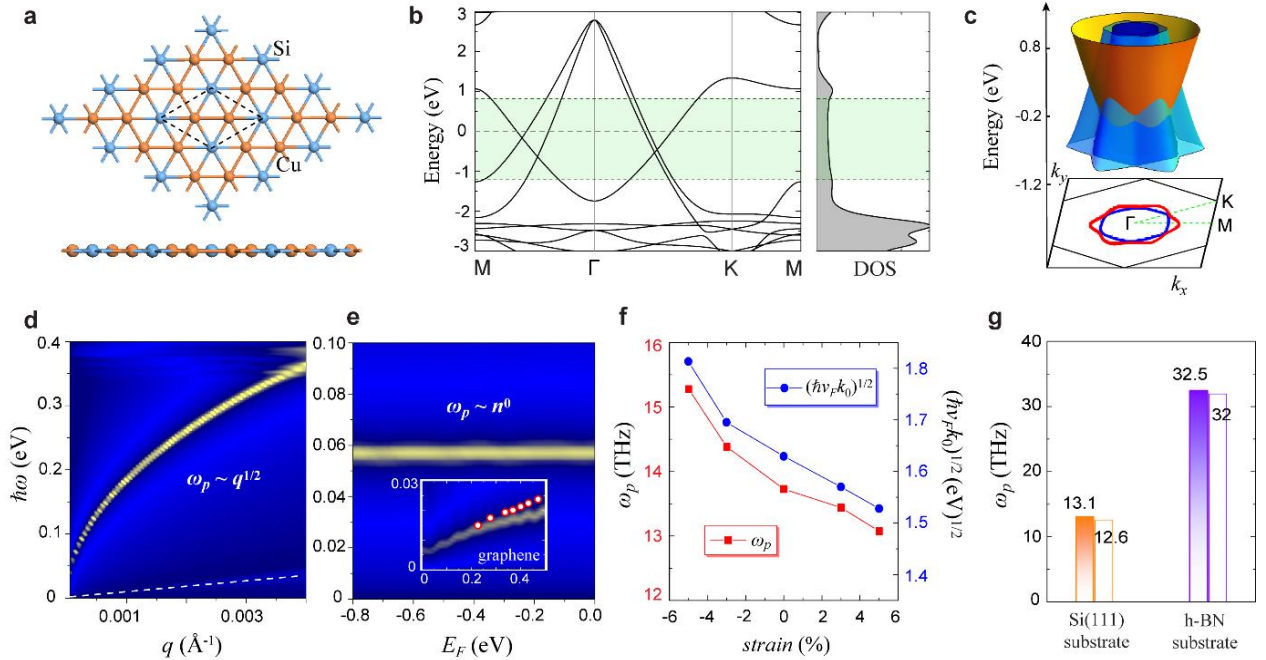

**Figure 2 | DIP in 2D NLSM Cu$_2$Si. a**, Top and side views of monolayer Cu$_2$Si. Black rhombus depicts primitive cell. **b**, Calculated band structure and DOS (without SOC) for Cu$_2$Si. Fermi level is set to zero. **c**, 3D plot of 2D bands in the energy range from −1.2 to 0.8 eV (green region in **b**). Band crossings between conduction and two valence bands are projected to the 2D BZ (red and blue loops). **d**, EELS as a function of frequency and wave vector. White-dashed line denotes the upper edge $\hbar v_F q$ of intraband particle-hole continuum. **e**, EELS as a function of frequency and Fermi energy ($q = 0.0001$ Å$^{-1}$). Inset: EELS of graphene plasmon with the same calculation parameters as Cu$_2$Si. Experimental data of graphene[14] are marked as red circles (their brightness indicates the plasmon oscillation strength). **f**, Calculated plasmon frequency and related physical quantities obtained from band structures under strain. **g**, Plasmon frequency with two different substrates. Solid columns: numerical results from first principles; hollow columns: analytical results calculated from Formula (3.7) in Table 1.

Next, we discuss the plasmon excitation of Cu$_2$Si. The dynamical dielectric function $\epsilon(\boldsymbol{q}, \omega)$ is numerically calculated with 2D Coulomb interaction $V(q)$ and a background dielectric screening of SiO$_2$/Si substrate (see Methods section). An electron energy loss spectrum (EELS) is given by the imaginary part of the inverse of $\epsilon(\boldsymbol{q}, \omega)$, whose broadened peaks indicate the plasmons[28]. As shown in Fig. 2d, a 2D plasmon dispersion ($\omega_p \sim q^{1/2}$) is demonstrated; similar to the case of graphene plasmons[40], it lies above the region of intraband electron-hole continuum, indicating that the direct Landau damping is forbidden. Using a typical micron wavelength ($q = 0.0001$ Å$^{-1}$), we plot the EELS as a function of $\omega_p$ and $E_F$ (corresponding to $n$) in Fig. 2e; it shows a clear $n$-independent feature ($\omega_p \sim n^0$), confirming the existence of DIP. As a comparison, the inset of Fig. 2e shows calculated results of graphene plasmon, agreeing well with the experimental data (red circles)[14]; the well-known $\omega_p \sim E_F^{1/2}$ ($\propto n^{1/4}$) relationship of graphene plasmon is revealed. In addition, the plasmon of Cu$_2$Si has a significantly larger oscillation strength than that of graphene, because of the higher DOS.

At a fixed micron wavelength, the plasmon of Cu$_2$Si stabilizes at a certain THz frequency with high intensity (Fig. 2e). As shown in Formula (3.7) in Table 1, the $\omega_p$ of 2D nodal line can be tuned by changing Fermi velocity $v_F$ and line node size $k_0$, which is demonstrated by the strain effect on Cu$_2$Si (Fig. 2f and Supplementary Fig. S2). The calculated $\omega_p$ and $(v_F k_0)^{1/2}$ have a consistent trend, confirming the validity of Formula (3.7). It notes that strain has a small effect on the change of $v_F$ and $k_0$; consequently, ~5% strain can slightly induce a ~2 THz change of $\omega_p$. On the other hand, $\omega_p$ is also sensitive to the background dielectric constant $\kappa$ [Formula (3.7)]. As shown in Fig. 2g, we further compare the $\omega_p$ of Cu$_2$Si on two substrates, i.e., Si(111) and h-BN, with quite different $\kappa$. Importantly, the nodal lines can survive on both substrates (see Supplementary Fig. S3). Interestingly, a 3-fold frequency change, from terahertz to mid-infrared, can be achieved by simply changing the underneath substrate of Cu$_2$Si.

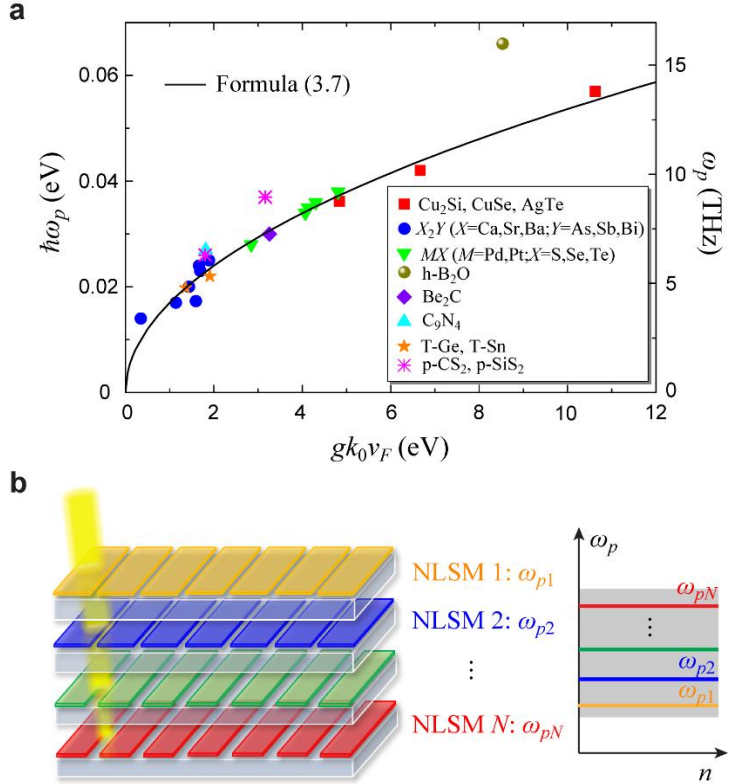

**Figure 3 | DIPs in 2D NLSM metamaterials for ultrastable and broadband terahertz spectroscopy. a**, Plasmon frequency as a function of $gk_0 v_F$ for 22 2D NLSMs, where $q = 0.0001$ Å$^{-1}$ and SiO$_2$/Si substrate are adopted (see Methods section). 22 colored points: numerical results obtained from first principles; black curve: analytical results obtained from Formula (3.7). **b**, Van der Waals heterostructures or grating superlattice of 2D NLSMs for ultrastable terahertz spectroscopy. For grating superlattice, transparent substrates are needed.

**A broadband terahertz spectroscopy in 2D NLSM metamaterials**

The intrinsic properties of nodal line (e.g., $v_F$ and $k_0$) are material-dependent, indicating that a broad range of DIPs with different $\omega_p$ may be achieved by the choice of different TSMs. Indeed, the DIP features are not only confirmed in all other NLSMs (see Supplementary Figs. S4−S11), but also exhibit a broad range from 4 THz (in $Ca_2As$) to 16 THz (in $h$-$B_2O$) (see Fig. 3a and Supplementary Table S2). Once again, the $\omega_p$ obtained from Formula (3.7) fit well with the first-principles results independent of their diverse lattice symmetries, except for the case of $h$-$B_2O$, which is due to the strong anisotropy of its nodal line[35] (see Supplementary Fig. S7).

Recently, the development of 2D van der Waals (vdW) heterostructures enables manipulating crystals for exploration of physics not observable in conventional materials[41,42]. Employing the heterostructures (or superlattice) of 2D NLSMs, here a novel ultrastable terahertz metamaterial device can be proposed, as drawn in Fig. 3b. For vdW stackings of 2D NLSMs, a compensation momentum is necessary for incident light, which can be realized in the scattering-type scanning near-field optical microscopy (s-SNOM) technology[16,27]. The 2D NLSMs can also be fabricated in micro-ribbon arrays on transparent substrates[14]; thus a grating superlattice can be well designed by stacking them layer by layer (Fig. 3b). Determined by the selections of fabricated NLSMs, ribbon/gap width and even substrates, multiple terahertz-frequency plasmons can be simultaneously excited. For instance, one can adopt the same or different materials/pattern periods for fabrications. Thus, a terahertz spectroscopy from narrowband to broadband can be achieved. Most importantly, the spectroscopy in such device could be ultrastable under the variable environments against charge doping. Our proposed device can be used as an ultrastable terahertz signal amplifier or an ultrastable terahertz sensor, which could selectively output or detect multi-band terahertz waves.

**Terahertz DIPs in 1D CNTs**

The nodal-point semimetals have been widely studied, *e.g.*, Dirac points in 2D graphene or 3D $Na_3Bi$[25]. To confirm our DIP model in 1D nodal-point systems [Formula (3.9) in Table 1], armchair CNTs have been selected as typical examples, as they are known as 1D Dirac-point semimetals[43].

The structure and 1D Dirac bands of armchair CNTs are shown in Supplementary Fig. S12. The calculations of (5,5), (10,10) and (15,15) nanotubes reveal that the $v_F$ of Dirac electrons are almost independent of their tube diameters $a$ and all of them have a constant DOS near the Fermi level (Supplementary Fig. S12). Using 1D Coulomb interaction and a BN substrate dielectric screening (see Methods section), the $\epsilon(\boldsymbol{q},\omega)$ and EELS of these CNTs can be obtained. As shown in Fig. 4a, the plasmon dispersion of (5,5) nanotube demonstrates a typical 1D feature ($\omega\sim|q|$ at long wavelength). It also lies above the region of intraband electron-hole continuum without direct Landau damping. With a micron wavelength ($q = 0.0001$ Å$^{-1}$), the EELS as a function of $\omega_p$ and $E_F$ (corresponding to $n$) for (5,5) and (10,10) nanotubes is calculated, as shown in Fig. 4b. Remarkably, the terahertz DIP feature ($\omega_p\sim n^0$) is revealed for both CNTs, and the $\omega_p$ calculated from first principles are consistent with the analytical results (red-dashed lines in Fig. 4b). Moreover, the $\omega_p$ of CNTs are almost $a$-independent, as a result of the weak $a$-dependent $v_F$, *i.e.*, there is only a weak logarithm dependence on the diameter of CNTs[44] [see Formula (3.9)]. Interestingly, the recent experimental observations on metallic nanotubes with different diameters confirm the existence of mid-infrared DIPs[27]. Our results are in good agreement with the experimental measurements (red triangles in Fig. 4a), also reflecting the weak $a$ dependence.

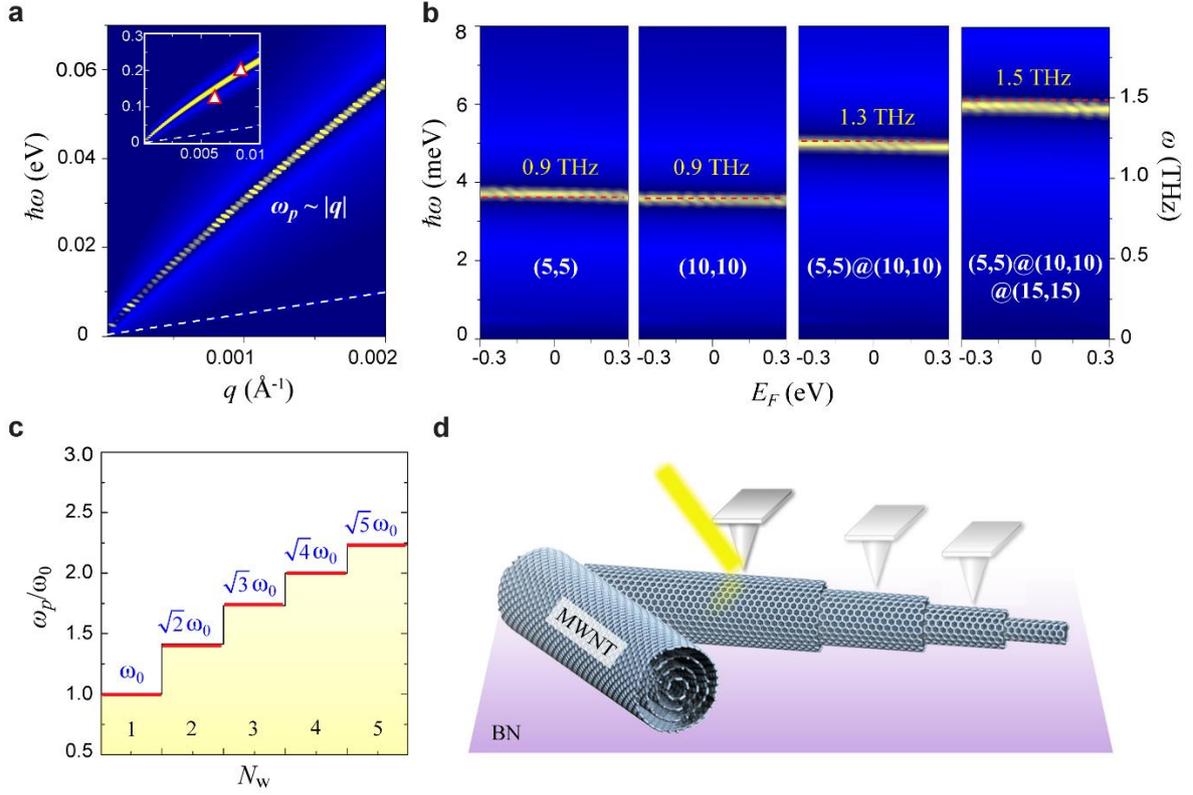

**Figure 4 | Terahertz DIPs in 1D metallic CNTs and their quantization. a**, EELS as a function of frequency and wave vector. White-dashed line denotes the upper edge $\hbar v_F q$ of intraband particle-hole continuum. Inset: EELS at a large scale of wave vector. Red triangles: experimental data adopted from ref. 27. **b**, EELS as a function of frequency and Fermi energy for two different single-, (5,5)@(10,10) double- and (5,5)@(10,10)@(15,15) triple-walled nanotubes ($q = 0.0001$ Å$^{-1}$). Red-dashed lines: analytical results from Formula (3.9) in Table 1. **c**, A quantized DIP frequency as a function of the number of walls ($N_w$) in multi-walled nanotubes (MWNT). **d**, Schematic plot of 1D MWNT metamaterial for an ultrastable terahertz spectroscopy with tunable and quantized frequencies.

**Quantized terahertz plasmons in 1D CNT metamaterials**

In terms of Formula (3.9), the weak $a$ dependence indicates that $\omega_p$ of CNTs is mostly determined by the degeneracy factor $g$. In multi-walled nanotubes (MWNT), the Dirac points could maintain due to the weak vdW interactions between the individual tubes (Supplementary Fig. S12); meanwhile, the long-range Coulomb interactions make conducting channels of electrons determined by the number of walls ($N_w$). Thus, we propose that a quantized terahertz $\omega_p$ may be achieved in MWNT metamaterials. As shown in Fig. 4b, the calculated plasmon excitations of (5,5)@(10,10) double-walled nanotubes and (5,5)@(10,10)@(15,15) triple-walled nanotubes are demonstrated. Interestingly, besides the DIP feature, the $\omega_p$ of MWNT can exhibit a clear quantized plateau as a function of $N_w$, as shown in Fig. 4c. It notes that a similar quantization of propagation velocity has been observed in single-walled CNT bundles[27], explained by a many-body Luttinger-liquid theory. Importantly, differing from the previous theory[27], the emergence of quantized $\omega_p$ comes naturally from our unified DIP theory in 1D system [Formula (3.9)]. Therefore, a quantized manipulation of ultrastable teraherz plasmons using a series of MWNT can be well designed (Fig. 4d), *e.g.*, discrete frequencies or wavelengths can be excited at different thickness of a telescoping MWNT. Meanwhile, a s-SNOM technology may be needed[16,27].

**Spatial confinement and lifetime of terahertz DIPs**

As two important figures of merit for plasmonics, spatial confinement and quality factor are also calculated for the DIPs in 2D NLSMs and 1D CNTs, in comparison to graphene plasmon (see Fig. 5 and Supplementary Fig. S13). The spatial confinement, defined as the ratio of free-space light wavelength and plasmon wavelength ($\lambda_0/\lambda_p$), is found to be related to the dimensions: with increasing the $\omega_p$, it can be enhanced for 2D systems but almost unchanged for 1D plasmons, consistent with the theoretical derivations (see Fig. 5a and Supplementary Section VII). Strong confinement effect is available for terahertz DIPs, *e.g.*, in (5,5) CNTs and Ca$_2$As (Fig. 5a), which is critical for the design of compact devices. The quality factor ($Q$), measuring the number of oscillating cycles a plasmon can propagate, is related to the plasmonic damping rate or lifetime. Due to the forbidden or weak direct Landau damping, the phonon-assisted damping rate of plasmons is solely considered (see Methods section). The calculated lifetimes ($\tau_p$) can reach tens of picoseconds at low $\omega_p$ but reduce rapidly when $\omega_p$ increases to the frequency of optical phonons (Fig. 5b).

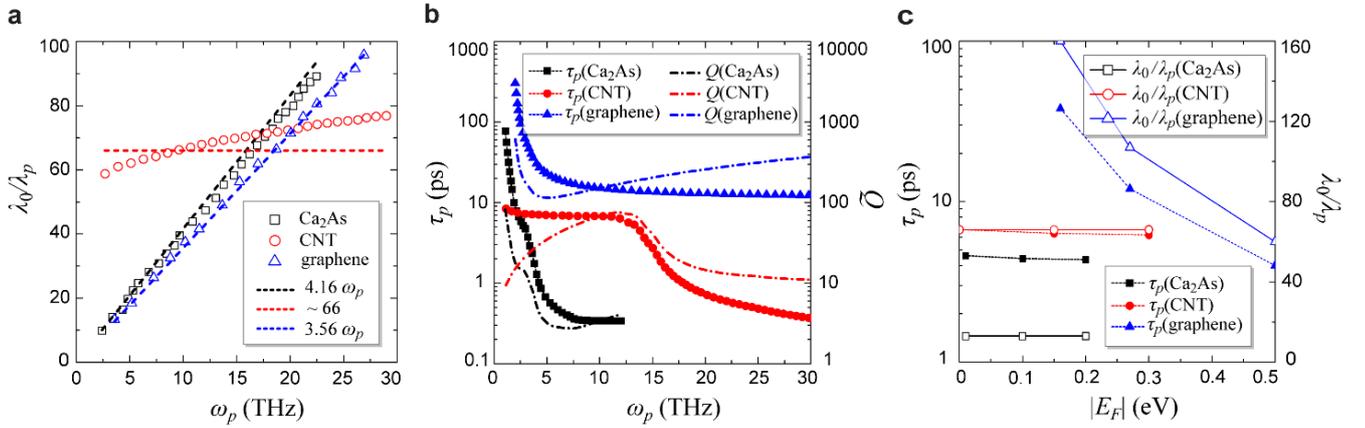

**Figure 5 | Spatial confinement and lifetime of DIPs. a,** Spatial confinement ($\lambda_0/\lambda_p$) as a function of $\omega_p$ for Ca$_2$As, (5,5) CNT and graphene. The dashed lines are theoretically derived $\lambda_0/\lambda_p$ for the three systems (see Supplementary Section VII). $E_F = 0.27$ eV is adopted for graphene, in order to compare with experiment[16]. **b,** Lifetime ($\tau_p$) and quality factor ($Q$) as functions of $\omega_p$ for Ca$_2$As, CNT and graphene. **c,** $\lambda_0/\lambda_p$ and $\tau_p$ as functions of $|E_F|$ for Ca$_2$As, CNT and graphene. The $\omega_p$ is fixed at 3, 10 and 30 THz for Ca$_2$As, CNT and graphene, respectively.

Usually, strong spatial confinement of a plasmon is achieved at the expense of a low quality factor for metals. Remarkably, the calculated mid-infrared graphene plasmon (e.g., at 30 THz) could simultaneously exhibit extraordinary spatial confinement ($\lambda_0/\lambda_p \sim 105$) and high quality factor ($Q \sim 360$), consistent with the experimental observations[16]. Interestingly, the simultaneous high spatial confinement and quality factor can be also achieved in terahertz DIPs, e.g., $\lambda_0/\lambda_p \sim 13$ (66), $Q \sim 15$ (67) for plasmon in Ca$_2$As (CNT) at 3 (10) THz. As shown in Fig. 5c, the performance ($\lambda_0/\lambda_p$ and $\tau_p$) of graphene plasmons can be greatly affected by the charge doping, which is also demonstrated in a recent experiment[18]. However, for terahertz DIPs, these figures of merit are robust against the change of *n* (Fig. 5c), revealing that the *n*-independent feature of DIPs also benefits their device performance.

**Discussion and Conclusion**

For 2D NLSMs, some of them have been synthesized on metal substrates[33,34,37]. Although the mirror reflection symmetry is broken considering the different media of vacuum and substrate, the nodal lines survive due to the weak substrate-overlayer interactions[37]. Because of the significantly different excitation regions of $\omega_p$ in these NLSMs (terahertz) and their metal substrates (ultraviolet), we expect that the novel DIPs are ready to be detected in the experiments. It is also expected that they can further be deposited on insulating substrates (such as $SiO_2$/Si or BN) by the transfer technique[42] for better optical measurements, where near-field optical microscope[16,27] or lithography and etching technologies[14] may need to be adopted. Meanwhile, some other 2D NLSMs are expected to be synthesized or exfoliated from their bulk materials. For 1D nodal-point TSMs, the large-scale, high-quality CNTs can be synthesized in the experiments[45].

In conclusion, we reveal a unified mechanism for generating unconventional DIPs in some specific topological systems, which can benefit the stable functionalities of plasmons in changeable environments especially for the stable terahertz applications. Our study has significant impacts in multiple fields: First, the unified mechanism we revealed not only could deepen our understandings on plasmons, but also provide a guide to the experimental realizations of unconventional DIPs in a series of toplogical matters. Second, combining with a large number of TSMs, a completely new type of terahertz plasmons are ready to be applied in compact terahertz devices with high stability and precision, such as ultrastable signal amplification and accurate detection. Especially, with the developments of vdW heterostructures[41,42], new concepts of terahertz topological metamaterials can be designed to achieve a terahertz-stable spectroscopy from narrowband to broadband or a quantized manipulation of terahertz frequencies. Therefore, our work paves the way to developing exotic plasmonic applications in nanophotonic and nanophotoelectric devices, which may potentially open a new field for terahertz-stable plasmonics and related technologies.

## Methods
### First-principles calculations

The first-principles calculations are performed using the Vienna ab initio simulation package[46] within the projector augmented wave method[47] and the generalized gradient approximation of the Perdew-Burke-Ernzerhof[48] exchange-correlation functional. The $\Gamma$-centered $k$-point meshes are adopted. Fixing the crystal symmetry, the structures from experiments or literatures are relaxed until the residual forces on each atom is less than 0.01 eV/Å. The thickness of vacuum is taken to be 18 Å, which is adequate to simulate 2D or 1D materials. $Cu_2Si$ has a lattice constant of 4.123 Å, and the crystal structures of other materials can be found in Supplementary Information. Spin-orbit coupling (SOC) is also considered in part of our calculations. A tight-binding (TB) Hamiltonian based on the maximally localized Wannier functions (MLWF)[49] is constructed to get the energy eigenvalues and eigenstates for further dielectric function calculations.

### Plasmon calculations

The plasmon excitation can be determined by
$$\epsilon(\boldsymbol{q},\omega) = 1 - V(q)\Pi(\boldsymbol{q},\omega) = 0, \qquad (4)$$
where $\epsilon(\boldsymbol{q},\omega)$, a function of the wave vector $\boldsymbol{q}$ and frequency $\omega$, is the dynamical dielectric function. $V(q)$ is the $D$-dimensional Coulomb interaction in the wave vector space[26]

$$V(q) = \begin{cases} 4\pi e^2/\kappa q^2, (D=3) \\ 2\pi e^2/\kappa q, (D=2) \\ 2e^2|\ln(qa)|/\kappa. (D=1) \end{cases} \qquad (5)$$

$\kappa = 4\pi\epsilon_0\epsilon_r$ is the background dielectric constant, where $\epsilon_0$ and $\epsilon_r$ are the vacuum and background relative dielectric constants, respectively. $\Pi(\boldsymbol{q},\omega)$ is the irreducible polarizability function. Under RPA and long-wavelength limit ($q \to 0$), the plasmon frequency in a $D$-dimensional electron system can be determined by a noninteracting irreducible polarizability[26,28]

$$\Pi(\boldsymbol{q},\omega) = \frac{g}{(2\pi)^D}\int d^D\boldsymbol{k} \sum_{l,l'} \frac{n_F(E_{\boldsymbol{k},l})-n_F(E_{\boldsymbol{k}+\boldsymbol{q},l'})}{\hbar\omega+E_{\boldsymbol{k},l}-E_{\boldsymbol{k}+\boldsymbol{q},l'}+i\eta} F_{ll'}(\boldsymbol{k},\boldsymbol{q}), \qquad (6)$$

in which $n_F$ is the Fermi-Dirac distribution function, and $F_{ll'}(\boldsymbol{k},\boldsymbol{q})$ is the overlap form factor $|\langle\boldsymbol{k}+\boldsymbol{q},l'|e^{i\boldsymbol{q}\cdot\boldsymbol{r}}|\boldsymbol{k},l\rangle|^2$, with $|\boldsymbol{k},l\rangle$ and $E_{\boldsymbol{k},l}$ the eigenstate and energy dispersion respectively. The factor $g$ in Eq. (6) is the degeneracy factor including degeneracies of spin, valley and conducting channel, and $\eta$ is related to the electron lifetime due to the damping. The zeros of complex dielectric function signify a self-sustaining collective mode and give the plasmon frequency.

**Numerical calculations:** The energy eigenvalues and eigenstates in Eq. (6) are obtained from the TB Hamiltonian of MLWF. The integral is over the first BZ, where a temperature of 300 K in the Fermi-Dirac distribution function and an infinitesimal broadening $\eta = 1$ meV are used. Considering the spin degree of freedom, the degeneracy factor $g$ is set to 2 in the numerical calculations of all systems; while in theoretically derived formula in Table 1, $g$ should be adopted as 4 for $Cu_2Si$, CuSe and AgTe, because of the two nodal lines in these three TSMs, and 4 (6) for double-walled nanotubes (triple-walled nanotubes). For $V(q)$, the relative dielectric constant is determined by $\epsilon_r = (\epsilon_0 + \epsilon_{\text{sub}})/2$ for 2D and 1D systems, representing the effective dielectric function of environments (vacuum and substrate). For a $SiO_2$/Si substrate, $\epsilon_r = 5$[14]; for a BN substrate, $\epsilon_r = 1$[27]. In the 1D Coulomb interaction, $a$ is lateral confinement size of 1D electron system, e.g., the diameter of nanotubes. When studying the density dependence of plasmon, we have fixed the wave vector as $q = 0.0001$ Å$^{-1}$, which corresponds to a typical micron wavelength easily available in experiments[14].

The collective plasmon mode is defined at zeros of Eq. (3). In general, the dielectric function is a complex functional. The complex solution at $\epsilon(\boldsymbol{q},\omega) = 0$ gives both the plasmon dispersion (real part) and the decay of the plasmon (imaginary part). In order to compare with the experiments, it is more convenient to calculate the electron energy loss spectrum (EELS), whose broadened peaks indicate the plasmons[28]

$$\text{EELS} = -\text{Im}[1/\epsilon(\boldsymbol{q},\omega)]. \qquad (7)$$

**Analytical derivations:** In the analytical derivations of $\omega_p$, only the intraband excitations are considered. $E(\boldsymbol{k}) = \hbar^2k^2/2m$ and $E(\boldsymbol{k}) = \hbar v_F k$ are employed for conventional EGs and TSMs, respectively. The overlap form factor is $F(\boldsymbol{k},\boldsymbol{q}) = 1$, and a Fermi-Dirac distribution function with 0 K is adopted. The detailed derivations for all systems are shown in Supplementary Information. The final results are summarized in the right column of Table 1. In comparison with the results of numerical calculations, the degeneracy factor $g$ should be adopted as 4 for $Cu_2Si$, CuSe and AgTe, because of the two nodal lines in these three TSMs, 4 (6) for double-walled nanotubes (triple-walled nanotubes), and 2 for other materials.

**Calculations of plasmon lifetime**

The phonon-assisted damping rate of plasmons is only considered, due to the forbidden or weak direct Landau damping in the studied systems at long-wavelength limit. The plasmon lifetime can be determined from the ac conductivity, which is related to the transport relaxation time ($\tau_{tr}$)[50]. The transport scattering rate can be written as[50]

$$\frac{1}{\tau_{tr}} = \frac{2\pi}{\omega}\int_0^\omega d\omega'(\omega-\omega')\alpha_{tr}^2 F(\omega'). \tag{8}$$

$\alpha_{tr}^2 F$ is transport-related Eliashberg spectral function and it is defined as

$$\alpha_{tr}^2 F(\omega) = \frac{1}{2N_F}\sum_{\boldsymbol{k}\boldsymbol{q}\nu}|g^\nu(\boldsymbol{k},\boldsymbol{q})|^2(-\frac{\boldsymbol{q}\cdot\boldsymbol{k}}{k^2})\delta(\varepsilon_{\boldsymbol{k}}-\varepsilon_F)\delta(\varepsilon_{\boldsymbol{k}+\boldsymbol{q}}-\varepsilon_F)\delta(\omega-\omega_{\boldsymbol{q}\nu}), \tag{9}$$

where $g^\nu(\boldsymbol{k},\boldsymbol{q})$ is the electron-phonon (*e-ph*) matrix element. It quantifies a scattering process from an initial Bloch state $|n\boldsymbol{k}\rangle$ (with band *n* and momentum *k*) to a final state $|m\boldsymbol{k}+\boldsymbol{q}\rangle$ by emitting or absorbing a phonon with wavevector *q*, mode index *ν* and frequency $\omega_{\nu\boldsymbol{q}}$,

$$g_{nm\nu}(\boldsymbol{k},\boldsymbol{q}) = \frac{1}{\sqrt{2\omega_{\nu q}}}\langle m\boldsymbol{k}+\boldsymbol{q}|\partial_{\boldsymbol{q}\nu}V|n\boldsymbol{k}\rangle, \tag{10}$$

where $\partial_{\boldsymbol{q}\nu}V$ is the derivative of the self-consistent potential.

The *e-ph* coupling matrix elements can be computed within the density-functional perturbation theory (DFPT). Here, a Wannier-Fourier interpolation method as implemented in the EPW code[51] and integrated in the Quantum ESPRESSO package[52] is used to obtain the numerical results of *e-ph* coupling. The electron eigenstates and eigenvalues, vibrational modes and frequencies, as well as *e-ph* matrix elements are first calculated on a relatively coarse BZ grid, and then Wannier-interpolated values on a fine grid are obtained.

**Acknowledgements**

The authors thank L. Kang for helpful discussions. J.W. and B.H. acknowledge the support from Science Challenge Project TZ2016003 and NSAF U1930402. X.S. and W.D. acknowledge support from MOST of China (Grant No. 2016YFA0301001), NSFC (Grants No. 11674188 and No. 11874035) and the Beijing Advanced Innovation Center for Future Chip (ICFC). F.L. acknowledge the support from US-DOE (Grant No. DE-FG02-04ER46148). Part of the calculations were performed at Tianhe2-JK at CSRC.

**Author contributions**